\documentclass{iopart}
\usepackage{graphics}
\newcommand{\rb}{{\bi{r}}}             

\newcommand{\bb}{{\bi{b}}}
\newcommand{\hb}{{\bi{h}}}

\newcommand{\Rb}{{\bi{R}}}
\newcommand{\Zeb}{{\bi{0}}}
\newcommand{\Ri}{{{\bi{R}}_i}}
\newcommand{\trho}{\tilde{\rho}}
\newcommand{\tF}{\tilde{F}}
\newcommand{\tU}{\tilde{U}}
\newcommand{\tW}{\tilde{W}}
\newcommand{\tRi}{{\tilde{\bi{R}_i}}}
\newcommand{\tEc}{\tilde{\mathcal{E}}}
\newcommand{\Ec}{{\mathcal{E}}}
\newcommand{\mOmega}{{\mathit{\Omega}}}
\newcommand{\erfc}{\mathop{\rm{erfc}}\nolimits}
\newcommand{\sump}{\mathop{{\sum}'}}

\newcommand{\intp}{\mathop{{\int}'}}
\begin{document}
\title[Multiple charge spreading as a generalization of the Bertaut 
approach]
{Multiple charge spreading as a generalization of the Bertaut 
approach to lattice summation of Coulomb series in crystals}
\author{Eugene V Kholopov}
\address{A V Nikolaev Institute of Inorganic Chemistry, Siberian Branch,
Russian Academy of Sciences, 630090 Novosibirsk, Russia}
\address{Novosibirsk State University, 630090 Novosibirsk, Russia}
\ead{kholopov@che.nsk.su}
\begin{abstract}
The Bertaut approach associated with charge spreading so as to enhance the rate
of convergence of Coulomb series in crystals is extended to the case of an
arbitrary multiple spreading with a given initial spreading function. It is 
shown that the effect of spreading may in general be treated as a uniform 
transformation of space, providing that zero mean potential as a universal 
spatial property is sustained. As a result, electrostatic potentials driven
by different orders of multiple spreading can be obtained from the same energy
functional in a consistent manner. It is found that the effect of multiple 
spreading gives rise to more advanced forms described, for example, by simple 
exponential decrease, but the functional description based on a Gaussian 
spreading turns out to be invariant. In addition, the effects of a multiple
charge spreading based on simple exponential and Gaussian spreading functions
are compared as typical of molecular calculations.
\end{abstract}

\section{Introduction}
The problem of lattice summation of Coulomb series over crystal structures
is principal for describing solid state. Apart from a large number of
traditional approaches to this subject \cite{Tosi64,Glas80,Khol04}, many novel 
proposals for solving this problem still arise \cite{Wolf99,Mars00,Demo01,%
Mars02,Venk02,Tyag04,Pask05,Tyag05,Harr06}. Nevertheless, the classical Ewald 
approach \cite{Ewal21} remains one of the most effective and so widespread 
\cite{Grzy00,Port00,Whee02,Zhan02}. This is the reason that understanding
the nature of this efficiency is of great importance \cite{Khol07}. In 
particular, its relation to the effect of screening Coulomb potentials was 
revealed by Nijboer and De Wette \cite{Nijb57}. As a result, some 
generalizations associated with different types of screening have been 
proposed \cite{Whee02,Sugi84}. Another fruitful explanation of the foregoing 
efficiency is based on the idea of charge spreading proposed first by Ewald
in his original paper \cite{Ewal21} and developed further by Bertaut 
\cite{Bert52}. In particular, this generalized approach turns out to be
expedient in applications to molecular dynamics \cite{Whee02,Luty95}. 
Here we will discuss this treatment in more amount of detail so as to 
coincide known variations inherent in its implementation.
 
In the original paper of Bertaut \cite{Bert52} the pair-wise Coulomb
interaction is discussed. As a result, the double charge spreading naturally
arises in that approach. In particular, this feature results in the fact
that the square of the Fourier transform describing the spreading function
takes place in the sum over the reciprocal lattice contributing to the Coulomb
energy \cite{Bert52,Temp55}. On the other hand, it turns out that a single 
charge spreading is still sufficient if the electrostatic potential is first 
considered \cite{Jenk71,Ween75,Herz81}. As a result, the Fourier transform 
of the spreading function, but not its square, arises in the sum over 
reciprocal lattice vectors contributing to the energy within such a treatment. 
This fact was the subject of discussion \cite{Ween75,Bert78,Argy92}. 
Altogether, it was shown that both the treatments are quite correct and can 
eventually originate the description proposed by Bertaut. Nevertheless, the 
original treatment proposed by Bertaut appears to be symmetric with respect 
to both the set of charges generating potentials and the similar set of 
charges interacting with those potentials. It is instructive that the latter 
situation may be regarded as some uniform transformation of space 
\cite{Khol04,Luty95}.

In the present paper we extend the concept of charge spreading and introduce 
the regular idea of a multiple spreading, bearing in mind that this effect 
can always be addressed to the charge distribution generating the potential 
field. On the other hand, such a standpoint is not obviously unique and 
therefore various other points of view are discussed. In particular, the idea 
of spreading as a uniform transformation of space is developed in a general 
form. In addition, in the case of a multiple spreading the universal 
character of the Ewald approach is recognized. The effect of a multiple
charge spreading on the Coulomb interaction between a couple of objects
neutral on average is also discussed.

\section{Preliminaries}
Let us consider a crystal described by a charge distribution
$\rho_{\rm{c}}(\rb)$ subject to translational symmetry. On the other hand, 
we can also introduce a local charge distribution $\rho(\rb)$ attributed 
to a unit cell and driven by the natural condition of electrical neutrality
\begin{equation}\label{Aq1}
\int_V\rho(\rb)\,d\rb=0 ,
\end{equation}
where the integration is carried out over a volume $V$ occupied by
$\rho(\rb)$. Note that the unit-cell parallelepiped of volume $v$,
constituted of three noncomplanar vectors of elementary translations of a
Bravais lattice at hand, is assumed to be contained completely in $V$.
The latter is essential if $\rho(\rb)$ is spread beyond that
parallelepiped volume \cite{Khol04}. Then the overall charge distribution
in question can be written as
\begin{equation}\label{Aq2}
\rho_{\rm{c}}(\rb)=\sum_i\rho(\rb-\Ri) ,
\end{equation}
where the summation over $i$ is performed over sites specified by vectors
$\Ri$ appropriate to the Bravais lattice of interest. It is evident
that representation (\ref{Aq2}) for $\rho_{\rm{c}}(\rb)$ is not unique
due to an optional choice of $\rho(\rb)$ \cite{Khol04}. Nevertheless,
form (\ref{Aq2}) is subject to translational symmetry as well. This is the
reason that $\rho_{\rm{c}}(\rb)$ defined by (\ref{Aq2}) can be cast in
terms of a series over reciprocal lattice vectors $\hb$:
\begin{equation}\label{Aq3}
\rho_{\rm{c}}(\rb)=\sump_\hb F(\hb)\exp(2\pi i \hb\rb) ,
\end{equation}
where the prime on the summation sign implies that the contribution of
$\hb=0$ is actually omitted, as follows from formula (\ref{Aq7}) derived 
later on. The structure factor $F(\hb)$, by definition, is determined as
\cite{Luty95,Argy92}
\begin{equation}\label{Aq4}
F(\hb)=\frac{1}{v}\int_{\rm{cell}}\rho_{\rm{c}}(\rb)\exp(-2\pi i \hb
\rb)\,d\rb .
\end{equation}
Here the integration is carried out over the unit-cell parallelepiped
mentioned above. Substituting (\ref{Aq2}) into (\ref{Aq4}), we obtain
\begin{equation}\label{Aq5}
F(\hb)=\frac{1}{v}\int_{\rm{cell}}d\rb\sum_i\rho(\rb-\Ri)
\exp(-2\pi i \hb(\rb-\Ri)) .
\end{equation}
Bearing in mind that the integration over the unit-cell parallelepiped
along with the summation over $i$ is transformed into the integration over
all space \cite{Beth28} reduced eventually to $V$ intrinsic to
$\rho(\rb)$, relation (\ref{Aq5}) is converted into
\begin{equation}\label{Aq6}
F(\hb)=\frac{1}{v}\int_V\rho(\rb)\exp(-2\pi i \hb\rb)\,d\rb .
\end{equation}
It is important that
\begin{equation}\label{Aq7}
F(\hb=0)=0 ,
\end{equation}
in agreement with (\ref{Aq1}).

The electrostatic potential exerted by charge distribution (\ref{Aq2}) at 
a reference point $\rb$ is of the form
\begin{equation}\label{Aq8}
U(\rb)=\intp\frac{\rho_{\rm{c}}(\rb_1)\,d\rb_1}{|\rb_1-\rb|} ,
\end{equation}
where the prime on the integral sign stands for missing a singular
contribution of any point charge if it happens at $\rb$. If we now
make use of Poisson's equation for the Green function
\begin{equation}\label{Aq9}
\nabla^2_\rb\frac{1}{|\rb_1-\rb|}=-4\pi\delta(\rb_1-\rb) ,
\end{equation}
where the differentiation is performed with respect to $\rb$ and 
$\delta(\rb)$ is the Dirac delta function, we readily confirm from
(\ref{Aq8}) that $U(\rb)$ is subject to the conventional Poisson's 
equation of the form 
\begin{equation}\label{Aq10}
\nabla^2_\rb U(\rb)=-4\pi\rho_{\rm{c}}(\rb).
\end{equation}

On the other hand, on inserting (\ref{Aq3}) into (\ref{Aq8}), the 
result can be written as
\begin{equation}\label{Aq11}
U(\rb)=\Bigl[\sump_\hb F(\hb)\int\frac{\exp(2\pi i\hb\rb_1)\,d\rb_1}{|
\rb_1-\rb|}\Bigr]' ,
\end{equation}
where the square brackets decorated by the prime imply missing the
same singular term mentioned in (\ref{Aq8}). It is easy to show that
\begin{equation}\label{Aq12}
\int\frac{\exp(2\pi i\hb\rb_1)\,d\rb_1}{|\rb_1|}=\frac{1}{\pi|\hb|^2} .
\end{equation}
Substituting (\ref{Aq12}) into (\ref{Aq11}), we obtain
\begin{equation}\label{Aq13}
U(\rb)=\frac{1}{\pi}\Bigl[\sump_\hb\frac{F(\hb)}{|\hb|^2}\exp(2\pi i
\hb\rb)\Bigr]' .
\end{equation}

However, if we insert relation (\ref{Aq2}) directly into 
(\ref{Aq8}), the result is as follows:
\begin{equation}\label{Aq14}
U(\rb)=\intp\frac{d\rb_1}{|\rb_1-\rb|}\sum_i\rho(\rb_1-\Ri) .
\end{equation}
If we go over to a new variable of integration $\rb'=\rb_1-\Ri$ now,
then
\begin{equation}\label{Aq15}
U(\rb)=\sump_i^*\int_V\frac{\rho(\rb')\,d\rb'}{|\tRi|} ,
\end{equation}
where
\begin{equation}\label{Aq16}
\tRi=\Ri+\rb'-\rb ,
\end{equation}
the prime on the summation sign in (\ref{Aq15}) means that the
singular contribution associated with a point charge at $\rb$ must be
still excluded. The asterisk over the summation sign points to the
fact that the summation over large $\Ri$ is not yet defined properly in
formula (\ref{Aq15}) so as to be consistent with the absence of the $\hb=0$
contribution in expression (\ref{Aq13}).

This inconsistency is the essence of the conditional convergence of Coulomb
series in crystals. It is especially pronounced in the particular case of
point-charge lattices described by \cite{Ween75}
\begin{equation}\label{Aq17}
\rho(\rb)=\sum_j q_j\delta(\rb-\bb_j) ,
\end{equation}
where the summation over $j$ is carried out over point charges $q_j$ 
belonging to a unit cell, located at positions $\bb_j$ and governed by
the condition
\begin{equation}\label{Aq18}
\sum_j q_j=0 ,
\end{equation}
in agreement with (\ref{Aq1}). Substituting (\ref{Aq17}) into
(\ref{Aq6}), we deduce
\begin{equation}\label{Aq19}
F(\hb)=\frac{1}{v}\sum_j q_j\exp(-2\pi i \hb\bb_j) .
\end{equation}
If relation (\ref{Aq19}) is now inserted into (\ref{Aq13}) and relation
(\ref{Aq17}) is inserted into (\ref{Aq15}), where equation (\ref{Aq16}) is
taken into account, then we obtain
\begin{equation}\label{Aq20}
U(\rb)=\frac{1}{\pi v}\Bigl\{\sump_{\hb,j}\frac{q_j}{|\hb|^2}
\exp[2\pi i\hb(\rb-\bb_j)]\Bigr\}'=\sump_i^*\sum_j\frac{q_j}{T_{ij}} ,
\end{equation}
where
\begin{equation}\label{Aq21}
T_{ij}=|\Ri+\bb_j-\rb| .
\end{equation}
Here the contribution of $\rb=\bb_j$ is supposed to be excluded in the
first relation on the right-hand side of (\ref{Aq20}) and the same 
contribution at $\Ri=0$ is to be excluded in the second issue. Both of 
the expressions in (\ref{Aq20}) describe the same potential by definition, 
so that the singularity associated with the contribution of large $\Ri$ 
must be resolved as it is prescribed by exclusion of the $\hb=0$ term. 
However, even in this case the convergence of the sum over $\hb$ 
is not fast for general $\rb$ \cite{Rein90}. The event of $\rb$ at which 
a point charge exists is an exclusion, where a compensating term enhancing 
the rate of convergence arises \cite{Harr70}.

\section{Multiple charge spreading}
In order to enhance the rate of convergence of the series mentioned above,
we extend the treatment of Bertaut \cite{Bert52} and define a modified
unit-cell charge distribution as follows:
\begin{eqnarray}\label{Bq1}
\trho^{(n)}(\rb)&=&\int\sigma(|\rb-\rb_1|)\sigma(|\rb_1-\rb_2|)\dots
\sigma(|\rb_{n-1}-\rb_n|)\nonumber\\
&&{}\times\rho(\rb_n)\,d\rb_1\dots d\rb_n ,
\end{eqnarray}
where $n$ identical functions $\sigma(|\rb|)$ are introduced. These
functions spread the actual charge at every point in a consecutive manner
and are normalized by the condition
\begin{equation}\label{Bq2}
\int\sigma(|\rb|)\,d\rb=4\pi\int_0^\infty\sigma(r)r^2\,dr=1 ,
\end{equation}
where $r=|\rb|$. Like (\ref{Aq2}), the modified overall charge distribution 
in the crystal then takes the form
\begin{equation}\label{Bq3}
\trho^{(n)}_{\rm{c}}(\rb)=\sum_i\trho^{(n)}(\rb-\Ri) .
\end{equation}
This value can in turn be cast in terms of the Fourier transforms
associated with the reciprocal lattice vectors:
\begin{eqnarray}
\trho^{(n)}_{\rm{c}}(\rb)=\sump_\hb \tF^{(n)}(\hb)\exp(2\pi i \hb\rb) ,
\label{Bq4}\\
\tF^{(n)}(\hb)=\frac{1}{v}\int_{\rm{cell}}\trho^{(n)}_{\rm{c}}(\rb_1)
\exp(-2\pi i \hb\rb_1)\,d\rb_1  .\label{Bq5}
\end{eqnarray}
Substituting (\ref{Bq3}) into (\ref{Bq5}), we obtain
\begin{eqnarray}\label{Bq6}
\tF^{(n)}(\hb)&=&\frac{1}{v}\int d\rb\exp(-2\pi i \hb\rb)
\int d\rb_1\dots d\rb_n\sigma(|\rb-\rb_1|)\nonumber\\
&&{}\times\sigma(|\rb_1-\rb_2|)\dots\sigma(|\rb_{n-1}-\rb_n|)\rho(\rb_n) .
\end{eqnarray}
If we here go over to new variables of integration
\begin{equation}\label{Bq7}
\rb'_{n-1}=\rb_{n-1}-\rb_n,\quad\dots,\quad
\rb'_1=\rb_1-\rb_2,\quad \rb'=\rb-\rb_1 ,
\end{equation}
keeping in mind that
\begin{equation}\label{Bq8}
\rb=\rb'+\rb'_1+\dots+\rb'_{n-1}+\rb_n ,
\end{equation}
then it is easy to show that
\begin{equation}\label{Bq9}
\tF^{(n)}(\hb)=F(\hb)S^n(\hb) .
\end{equation}
Here the function $S(\hb)$ is defined by the relation
\begin{equation}\label{Bq10}
S(\hb)=\int\sigma(|\rb|)\exp(-2\pi i \hb\rb)\,d\rb ,
\end{equation}
with the evident properties
\begin{equation}\label{Bq11}
S(\hb)=S(-\hb),\qquad S(0)=1 ,
\end{equation}
in agreement with formula (\ref{Bq2}).

The modified electrostatic potential appropriate to (\ref{Bq3})
is naturally equal to
\begin{equation}\label{Bq12}
\tU^{(n)}(\rb)=\int\frac{\trho_{\rm{c}}^{(n)}(\rb_1)\,d\rb_1}{|\rb_1
-\rb|} .
\end{equation}
Indeed, the substitution of (\ref{Aq9}) into (\ref{Bq12}) yields
Poisson's equation
\begin{equation}\label{Bq13}
\nabla^2_\rb \tU^{(n)}(\rb)=-4\pi\trho_{\rm{c}}^{(n)}(\rb_1)
\end{equation}
associated with (\ref{Bq1}). Comparing (\ref{Bq12}) with (\ref{Aq8}) 
and taking relations (\ref{Aq13}) and (\ref{Bq9}) into account, we 
readily derive
\begin{equation}\label{Bq14}
\tU^{(n)}(\rb)=\frac{1}{\pi}\sump_\hb\frac{F(\hb)S^n(\hb)}{|\hb|^2}
\exp(2\pi i\hb\rb) ,
\end{equation}
where any restriction associated with a point charge contribution is
immaterial now due to the attenuation effect of $S(\hb)$. This is a direct
consequence of the fact that there are no point charges after
transformation (\ref{Bq1}).

On the other hand, if we insert definition (\ref{Bq1}) into (\ref{Bq12})
and make use of relations (\ref{Bq7}) and (\ref{Bq8}), then we obtain
\begin{equation}\label{Bq15}
\tU^{(n)}(\rb)=\sum_i^*\int_V\rho(\rb')\mOmega^{(n)}(|\tRi|)\,d\rb' ,
\end{equation}
where definition (\ref{Aq16}) is utilized, the asterisk over the summation
sign points out that the problem of remote $\Ri$ still exists in the
present representation,
\begin{equation}\label{Bq16}
\mOmega^{(n)}(|\Rb|)=\int\frac{\sigma(|\rb_1|)\dots\sigma(|\rb_n|)\,
d\rb_1\dots d\rb_n}{|\Rb+\rb_1+\dots+\rb_n|} .
\end{equation}

To overcome the problem of remote $\Ri$, the initial electrostatic 
potential of interest can be rewritten in the form
\begin{equation}\label{Bq17}
U_{(n)}(\rb)\equiv\tU^{(n)}(\rb)+\Bigl[U(\rb)-\tU^{(n)}(\rb)\Bigr] .
\end{equation}
If we now utilize result (\ref{Bq14}) for the first term on the right-hand
side of formula (\ref{Bq17}) and employ (\ref{Aq15}), (\ref{Aq16}) and 
(\ref{Bq15}) for the remainder, we get
\begin{eqnarray}\label{Bq18}
U_{(n)}(\rb)&=&\frac{1}{\pi}\sump_\hb\frac{F(\hb)S^n(\hb)}{|\hb|^2}
\exp(2\pi i\hb\rb)\nonumber\\
&&{}+\sump_i\int_Vd\rb'\rho(\rb')\frac{W^{(n)}(|\tRi|)}{|\tRi|}
-\Bigl\{q_j\mOmega^{(n)}(0)\Bigr\}_{\rb=\bb_j} .
\end{eqnarray}
Here we introduce the following compact definition 
\begin{equation}\label{Bq19}
\frac{W^{(n)}(R)}{R}=\frac{1}{R}-\mOmega^{(n)}(R)
\end{equation}
for the difference characteristic of the case. As a result, the asterisk 
over the summation sign can be omitted, because the summation over $i$ in 
(\ref{Bq18}) is now carried out in a consistent manner resolving the 
conditional convergence of this sum at large $\Ri$. On the other hand, the 
prime on the summation sign over $i$ in (\ref{Bq18}) stands for the omission 
of the singular contribution of a point charge, if it happens, provided that 
such a contribution would be described by the first term on the right-hand 
side of (\ref{Bq19}). Finally, the last term on the right-hand side of
(\ref{Bq18}) describes the elimination of the same contribution from the
regular part specified by $\mOmega^{(n)}(R)$ in (\ref{Bq19}).

It is important that the convergence of the first term on the right-hand 
side of expression (\ref{Bq18}) is expected to be rather fast due to the 
effect of $S(\hb)$ and the same is right for the direct sum in the
remainder, in accord with \cite{Bert52}, as will be discussed in more
amount of detail in \cite{Kho208}.

Note that in the case of $n=1$ result (\ref{Bq18}) is tantamount to the 
Nijboer-De Wette approach \cite{Nijb57}, bearing in mind that 
normalization (\ref{Bq2}) is not principal here due to the fact that this 
representation is after all incorporated by means of identity (\ref{Bq17}).

\section{Charge spreading as a uniform transformation of space}
It is important that the spreading at hand can be treated as a uniform 
transformation of space \cite{Khol04}. Indeed, according to (\ref{Bq1}), 
this transformation connecting an initial point $\rb'$ with a final point 
$\rb$ is of the form
\begin{eqnarray}\label{Cq1}
f^{(n)}(\rb,\rb')&=&\int\sigma(|\rb-\rb_1|)\sigma(|\rb_1-\rb_2|)\dots
\sigma(|\rb_{(n-2)}-\rb_{(n-1)}|)\nonumber\\
&&{}\times\sigma(|\rb_{n-1}-\rb'|)\,d\rb_1\dots d\rb_{n-1} ,
\end{eqnarray}
where the limiting cases of $n=0$ and $n=1$ can be defined, respectively, 
as
\begin{eqnarray}
f^{(0)}(\rb,\rb')=\delta(\rb-\rb') ,\label{Cq2}\\
f^{(1)}(\rb,\rb')=\sigma(|\rb-\rb'|) .\label{Cq3}
\end{eqnarray}
Note that in terms of (\ref{Cq1}), definition (\ref{Bq1}) takes the form 
\begin{equation}\label{Cq4}
\trho^{(n)}(\rb)=\int f^{(n)}(\rb,\rb')\rho(\rb')\,d\rb' .
\end{equation}
According to (\ref{Bq2}), one can see that
\begin{equation}\label{Cq5}
\int f^{(n)}(\rb,\rb')\,d\rb'=1 .
\end{equation}
Moreover, it is evident from definition (\ref{Cq1}) that
\begin{equation}\label{Cq6}
f^{(n)}(\rb,\rb')=f^{(n)}(\rb',\rb) .
\end{equation}
This symmetry implies that transformation (\ref{Cq1}) may be regarded
either as a spreading of initial points containing charges or as a
spreading of final points which may be free from charges. 
Furthermore, relation (\ref{Cq1}) can be represented as a following
convolution:
\begin{equation}\label{Cq7}
f^{(n)}(\rb,\rb')=\int f^{(m)}(\rb,\rb_1)f^{(n-m)}(\rb_1,\rb')
d\rb_1 ,
\end{equation}
where $0\leq m\leq n$, with including the limiting cases specified by
(\ref{Cq2}) and (\ref{Cq3}) in the integrand. 

Another important convolution arises from (\ref{Bq16}) as connecting the 
initial point $\rb'$ and the final point $\rb$ in (\ref{Aq16}). Indeed, 
we can substitute expression (\ref{Aq16}) in place of $\Rb$ in formula 
(\ref{Bq16}) and go over to the following new variables of integration 
$\rb'_j$:  
\begin{eqnarray}
\rb_1=\rb-\rb_1',\quad \rb_2=\rb'_1-\rb'_2,\;\;\dots,\;\; \rb_m=\rb'_{m-1}-
\rb'_m,\label{Cq8}\\
\rb_n=\rb'_n-\rb',\quad\rb_{n-1}=\rb'_{n-1}-\rb'_n,\dots,\;\; \rb_{m+1}=
\rb'_{m+1}-\rb'_{m+2},\label{Cq9}
\end{eqnarray}
where $0\leq m\leq n$ again. Integrating over $\rb'_j$ and keeping 
relation (\ref{Cq1}) in mind, we get
\begin{equation}\label{Cq10}
\mOmega^{(n)}(|\Ri+\rb'-\rb|)=\int\frac{f^{(m)}(\rb,\rb_1)f^{(n-m)}
(\rb_2,\rb')\,d\rb_1\,d\rb_2}{|\Ri+\rb_2-\rb_1|} .
\end{equation}

Inserting (\ref{Cq7}) into (\ref{Cq4}) and (\ref{Cq10}) into 
(\ref{Bq15}), one can readily show that 
\begin{eqnarray}
\trho^{(n)}(\rb)=\int f^{(m)}(\rb,\rb')\trho^{(n-m)}(\rb')\,d\rb' ,
\label{Cq11}\\
\tU^{(n)}(\rb)=\int f^{(m)}(\rb,\rb')\tU^{(n-m)}(\rb')\,d\rb' ,
\label{Cq12}
\end{eqnarray}
where $0\leq m\leq n$ and definitions (\ref{Cq4}) and (\ref{Bq15}) are,
respectively, used in the integrands. Note that relations (\ref{Cq11}) 
and (\ref{Cq12}) are of the same structure. Moreover, they turn out to be
complementary to each other. The latter fact becomes evident if we
consider the bulk Coulomb energy per unit cell, which can be written down
in a traditional fashion \cite{Ween75,Argy92,Khol06} as
\begin{equation}\label{Cq13}
\Ec_{(n)}=\frac{1}{2}\int_V\rho(\rb)U_{(n)}(\rb)\,d\rb .
\end{equation}
On substituting (\ref{Bq18}) into (\ref{Cq13}) and taking equation 
(\ref{Aq6}) into account, relation (\ref{Cq13}) is easily converted into
\begin{eqnarray}\label{Cq14}
\Ec_{(n)}&=&\frac{v}{2\pi}\sump_\hb\frac{|F(\hb)|^2S^n(\hb)}{|\hb|^2}
+\frac{1}{2}\sump_i\int_Vd\rb\,d\rb'\rho(\rb)\rho(\rb')\nonumber\\
&&{}\times\frac{W^{(n)}(|\tRi|)}{|\tRi|}-\frac{\mOmega^{(n)}(0)}{2}
\sum_jq_j^2 ,
\end{eqnarray}
where the last term describes the correcting contribution of all point
charges in the unit cell. Upon investigating the first term on the
right-hand side, we may notice that the numerator of the summand can
be represented in the form 
\begin{equation}\label{Cq15}
|F(\hb)|^2S^n(\hb)=\tF^{(m)}(\hb)\tF^{(n-m)*}(\hb) ,
\end{equation}
where $0\leq m\leq n$, in agreement with (\ref{Bq9}). In other words, the
effect of spreading can be distributed between the couple of structure 
factors in an arbitrary manner.

Likewise, the temporary energy $\tEc^{(n)}$ associated with
$\tU^{(n)}(\rb)$ and contributing to (\ref{Cq14}) can be presented in
the form
\begin{equation}\label{Cq16}
\tEc^{(n)}=\frac{1}{2}\int_V\rho(\rb)\tU^{(n)}(\rb)\,d\rb .
\end{equation}
According to (\ref{Cq11}) and (\ref{Cq12}), one can see that formula 
(\ref{Cq16}) can also be rewritten as
\begin{equation}\label{Cq17}
\tEc^{(n)}=\frac{1}{2}\int\trho^{(m)}(\rb)\tU^{(n-m)}(\rb)\,d\rb 
\end{equation}
at $0\leq m\leq n$. This fact justifies the complementary character
of results (\ref{Cq11}) and (\ref{Cq12}). 

In terms of the space transformation it implies that two charged examples 
of transformed space interact either via (\ref{Cq17}) or via the first
term on the right-hand side of (\ref{Cq14}) with account of (\ref{Cq15}). 
In the symmetric case of $n=2m$ both of these examples of space appear to 
be identical. From the standpoint of symmetry, such an event of the highest 
symmetry relative to the effect of spreading is the most beautiful. The 
corresponding symmetric case at $n=2$ is the essence of the original 
treatment of Bertaut \cite{Bert52,Bert78}. 

Nevertheless, the chief objective of spreading is to improve the calculation 
of electrostatic potentials in crystals. This is the reason that all the 
effect of spreading, notwithstanding is it single or multiple, should 
practically be attributed to the potential part of the Coulomb energy that 
is eventually in conjunction with the principal idea of Bertaut 
\cite{Bert78,Argy92}. 

\section{General properties of bulk potentials at a multiple spreading}
The close connection between the Coulomb energy and electrostatic potentials
results in the known fact that the potential at any point can be determined
as a variational derivative of the energy at hand with respect to the charge
density at the same point \cite{Khol06}. With making use of relation 
(\ref{Cq13}), it implies that
\begin{equation}\label{Gq1}
\frac{\delta\Ec_{(n)}}{\delta\rho(\rb)}=U_{(n)}(\rb) ,
\end{equation}
keeping in mind that this result is quite general and so it is numerically 
independent of the subscript $n$, as mentioned above. 

A similar result but with a distinct interpretation appears if we deal with 
the energy determined by formula (\ref{Cq17}). In this case the revised  
version of (\ref{Gq1}) takes the form
\begin{equation}\label{Gq2}
\frac{\delta\tEc^{(n)}}{\delta\trho^{(m)}(\rb)}=\tU^{(n-m)}(\rb) ,
\end{equation}
where the restriction $0\leq m\leq n$ means that there are $n+1$ different 
events associated with definition (\ref{Gq2}). In other words, we have 
derived that the potential fields $\tU^{(n-m)}(\rb)$ with different 
superscripts can arise from a given $\tEc^{(n)}$. Note that along with the 
$n$th power of $S(\hb)$ in (\ref{Bq18}), this ambiguity for $n=2$ was 
discussed earlier \cite{Luty95}. For completeness, it should be emphasized 
that the same potential field $\tU^{(n-m)}(\rb)$ can be also obtained from 
$\tEc^{(n)}$ corresponding to different $n$. To this end, formula (\ref{Gq2}) 
has to be rewritten as follows:
\begin{equation}\label{Gq3}
\frac{\delta\tEc^{(n+k)}}{\delta\trho^{(m+k)}(\rb)}=\tU^{(n-m)}(\rb) ,
\end{equation}
where $k\geq-m$. Thus, issue (\ref{Gq2}) may be treated as a particular case 
of (\ref{Gq3}) at $k=0$. Relations (\ref{Gq2}) and (\ref{Gq3}) enable one 
to render some debatable places associated with charge spreading more tractable.
Indeed, according to (\ref{Bq13}), each of the potentials occurring in 
(\ref{Gq2}) and (\ref{Gq3}) corresponds to the solution of Poisson's equation 
specified by the charge distribution $\trho^{(n-m)}(\rb)$ appropriate to 
the case. In this respect, these potentials are quite determinate. On the other 
hand, the connection between these potentials and the energies specified
by (\ref{Cq17}) is also definite, despite the fact that different energies 
associated with the effect of charge spreading can be built up on the ground 
of the same potential field. This inference agrees with the conclusion known 
in the literature \cite{Luty95,Bert78,Argy92}.

Interested in general spatial properties of potentials connected with the 
charge spreading, now we discuss the mean potential value defined as:
\begin{equation}\label{Gq4}
\bar{U}=\frac{1}{v}\int_{\rm{cell}}U(\rb)\,d\rb .
\end{equation}
Substituting the first term on the right-hand side of (\ref{Bq18}) into
(\ref{Gq4}), we encounter with the relation
\begin{equation}\label{Gq5}
\frac{1}{v}\int_{\rm{cell}}\exp(2\pi i\hb\rb)\,d\rb=\delta_{\hb0} ,
\end{equation}
where $\delta_{\hb0}$ is the Kronecker delta. Formula (\ref{Gq5}) is the
fundamental relation of orthogonality describing the transformation from
the real space representation to the reciprocal space one. Indeed, if (\ref{Aq3}) 
is substituted into (\ref{Aq4}), the result becomes the identity due to 
relation (\ref{Gq5}). Hence, one can see that owing to the absence of the 
$\hb=0$ contribution to the first term on the right-hand side of (\ref{Bq18}), 
this contribution does not affect the value of (\ref{Gq4}).

Considering the contribution of the second term on the right-hand side 
of (\ref{Bq18}) to (\ref{Gq4}), we focus on the relation
\begin{equation}\label{Gq6}
G^{(n)}=\frac{1}{v}\int_{\rm{cell}}d\rb\sum_i\Bigl[\frac{1}{|\tRi|}
-\mOmega^{(n)}(|\tRi|)\Bigr]
\end{equation}
appearing in this case. Similar to the transformation from (\ref{Aq5})
to (\ref{Aq6}), the integration over the unit cell along with the
summation over $i$ is transformed to the integration over all space again.
As a result, expression (\ref{Gq6}) takes the form
\begin{equation}\label{Gq7}
G^{(n)}=\frac{1}{v}\int d\rb\Bigl[\frac{1}{r}-\mOmega^{(n)}(|\rb|)\Bigr] ,
\end{equation}
where definition (\ref{Aq16}) is taken into account and the corresponding
shift $\rb\to\rb-\rb'$ is suggested without changing the result. The
integration over the angular variables of $\rb$ in the first term in the
square brackets is trivial, whereas in the second one it is readily 
performed if we go over to the new variabe defined by equation (\ref{Zq2}) 
in \ref{App1}. Then we get
\begin{eqnarray}\label{Gq8}
G^{(n)}&=&\frac{4\pi}{v}\Bigl[\int_0^\infty rdr-\int\sigma(|\rb_1|)
\dots\sigma(|\rb_n|)\,d\rb_1\dots d\rb_n\nonumber\\
&&{}\times\Bigl(\frac{1}{Q}\int_0^Qr^2dr+\int_Q^\infty r\,dr\Bigr)\Bigr] ,
\end{eqnarray}
where $Q=|\rb_1+\dots+\rb_n|$. If the first term in the square brackets in
(\ref{Gq8}) is formally multiplied by $n$ integrals of form (\ref{Bq2}), 
then it can be combined with the second term therein. The result is as 
follows:
\begin{equation}\label{Gq9}
G^{(n)}=\frac{4\pi}{v}\int\sigma(|\rb_1|)\dots\sigma(|\rb_n|)
d\rb_1\dots d\rb_n\int_0^Q\Bigl(1-\frac{r}{Q}\Bigr)r\,dr .
\end{equation}
The integration over $r$ is straightforward here and we obtain
\begin{equation}\label{Gq10}
G^{(n)}=\frac{2\pi}{3v}\int Q^2\sigma(|\rb_1|)\dots\sigma(|\rb_n|)
d\rb_1\dots d\rb_n .
\end{equation}
It is significant that
\begin{equation}\label{Gq11}
Q^2=r_1^2+\dots+r_n^2+2(\rb_1\rb_2)+\dots+2(\rb_{n-1}\rb_n) ,
\end{equation}
where all scalar products vanish after integrating over angles in 
(\ref{Gq10}), but all terms $r_j^2$ give equal contributions to 
(\ref{Gq10}). Keeping relation (\ref{Bq2}) in mind, we finally obtain
\begin{equation}\label{Gq12}
G^{(n)}=\frac{2\pi n}{3v}\int r^2\sigma(r)\,d\rb
=\frac{8\pi^2n}{3v}\int_0^\infty r^4\sigma(r)\,dr .
\end{equation}
It is not surprising that result (\ref{Gq12}) looks like the mean potential
of Bethe \cite{Beth28} addressed to $n$ 'charge' distributions $\sigma(r)$
in a unit cell. It is important that $G^{(n)}$ turns out to be a constant.
Thus after substituting (\ref{Gq6}) into formula (\ref{Bq18}), result
(\ref{Aq1}) arises and so this contribution is zero as well. The last term
on the right-hand side of (\ref{Bq18}) is defined on a set of discrete 
points. Therefore its contribution to (\ref{Gq4}) is of measure zero 
and so it is negligible. As a consequence, we deduce
\begin{equation}\label{Gq13}
\bar{U}=0 .
\end{equation}
In other words, in uniform space zero mean charge, even with the effect
of spreading, generates zero mean potential \cite{Khol04,Khol06}.

\section{Simple exponential spreading}
For practical calculations some special representation of issue (\ref{Bq16})
is of interest. Indeed, as shown in \ref{App1}, formula (\ref{Bq16}) 
can be rewritten in the following recursion form
\begin{equation}\label{Dq1}
\mOmega^{(n)}(R)=\frac{2\pi}{R}\int_0^\infty\sigma(r)r\,dr
\int_{|R-r|}^{R+r}\mOmega^{(n-1)}(y)y\,dy ,
\end{equation}
where $\mOmega^{(0)}(R)=1/R$. Based on this relation, one can also
obtain the limiting result useful in what follows:
\begin{equation}\label{Dq2}
\mOmega^{(n)}(0)=4\pi\int_0^\infty\sigma(r)\mOmega^{(n-1)}(r)r^2dr .
\end{equation}

In the particular case of $n=1$ the values of $W^{(1)}(\Rb)$ and
$\mOmega^{(1)}(0)$ follow from (\ref{Bq19}) and (\ref{Dq1}) and from 
(\ref{Dq2}), respectively:
\begin{eqnarray}
W^{(1)}(R)=4\pi\int_R^\infty\sigma(r)r\bigl(r-R\bigr)dr ,\label{Dq3}\\
\mOmega^{(1)}(0)=4\pi\int_0^\infty\sigma(r)r\,dr ,\label{Dq4}
\end{eqnarray}
where equation (\ref{Bq2}) is employed. 

The values of $W^{(2)}(R)$ and $\mOmega^{(2)}(0)$ are also of
special interest. Their calculation is more tedious and is represented in
\ref{App1}. The corresponding general results are as follows: 
\begin{eqnarray}
W^{(2)}(R)&=&4\pi^2\Bigl[\int_0^\infty
dr\int_0^\infty dr'\!A(r,r')-\int_0^Rdr\int_0^{R-r} dr'\!A(r,r')\nonumber\\
&&{}-2\int_0^\infty dr\int_{R+r}^\infty dr'B(r,r')\Bigr] ,\label{Dq5}\\
\mOmega^{(2)}(0)&=&32\pi^2\int_0^\infty\sigma(r)r\,dr
\int_0^r\sigma(r')(r')^2\,dr' ,\label{Dq6}
\end{eqnarray}
where in formula (\ref{Dq5}) we introduce the notations:
\begin{eqnarray}
A(r,r')=\sigma(r)\sigma(r')rr'(R-r-r')^2 ,\label{Dq7}\\
B(r,r')=\sigma(r)\sigma(r')rr'(R+r-r')^2 .\label{Dq8}
\end{eqnarray}
It is worth noting that $\mOmega^{(2)}(R)$ determined by (\ref{Bq16})
and associated with $W^{(2)}(R)$ in form (\ref{Dq5}) through 
(\ref{Bq19}), may be regarded as the energy of Coulomb interaction 
between two 'charge' distributions $\sigma(r)$ of the distance $R$ apart 
\cite{Ween75}. Likewise, $\mOmega^{(2)}(0)$ is appropriate to the energy 
of self-interaction, in agreement with the Bertaut treatment 
\cite{Bert52,Ween75}.

There is a large variety of spreading functions discussed in the
literature \cite{Ewal21,Whee02,Bert52,Luty95,Ween75,Herz81,Bert78,%
Argy92,Kana55,Jone56,Herz79,Heye81}. Interested in principal aspects of
charge spreading, here we first consider the simplest spreading function 
$\sigma(r)$ which falls off exponentially with the distance $r=|\rb|$ 
\cite{Heye81,Birm58}:
\begin{equation}\label{Dq9}
\sigma(r)=\frac{\alpha^3}{8\pi}\exp\bigl(-\alpha r\bigr) ,
\end{equation}
providing that this function is normalized in compliance with (\ref{Bq2}). 
Substituting (\ref{Dq9}) into (\ref{Bq10}), we readily derive
\begin{equation}\label{Dq10}
S(\hb)=\Bigl[1+\Bigl(\frac{2\pi|\hb|}{\alpha}\Bigr)^2\Bigr]^{-2} .
\end{equation}
As far as $W^{(n)}(R)$ is concerned, we notice that this value is
dimensionless in accord with its definition (\ref{Bq19}). It is then
evident that this value can be cast in the form
\begin{equation}\label{Dq11}
W^{(n)}(R)=\tW^{(n)}(z) ,
\end{equation}
where $z$ is the dimensionless combination of $R$ and the spreading
parameter. In the present case it implies that $z=\alpha R$. After 
inserting (\ref{Dq9}) into (\ref{Dq3}) and (\ref{Dq4}), we obtain
\begin{eqnarray}
&&\tW^{(1)}(z)=\Bigl(1+\frac{z}{2}\Bigr)\exp(-z) ,\label{Dq12}\\
&&\mOmega^{(1)}(0)=\frac{\alpha}{2} ,\label{Dq13}
\end{eqnarray}
where result (\ref{Dq13}) follows from the combination of (\ref{Bq19}) 
and (\ref{Dq12}) as $R$ tends to zero. The case 
of $n=1$ arises upon substituting formulae (\ref{Dq10})--(\ref{Dq13}) 
into (\ref{Bq18}).

If the multiple spreading associated with $n=2$ is concerned, result 
(\ref{Dq10}) is still suitable. Substituting (\ref{Dq9}) into expression 
(\ref{Dq5}), we in turn obtain
\begin{equation}\label{Dq14}
\tW^{(2)}(z)=\Bigl(1+\frac{11z}{16}+\frac{3z^2}{16}+\frac{z^3}{48}\Bigr)
\exp(-z) .
\end{equation}
The value of 
\begin{equation}\label{Dq15}
\mOmega^{(2)}(R)=\frac{1}{R}-\Bigl(\frac{1}{R}+\frac{11\alpha}{16}+
\frac{3\alpha^2R}{16}+\frac{\alpha^3R^2}{48}\Bigr)\exp(-\alpha R) .
\end{equation}
is reconstructed from (\ref{Dq14}) with the help of (\ref{Bq19}) and 
(\ref{Dq14}). It is important that formula (\ref{Dq15}) describes the 
interaction between two identical exponential charge distributions of
which centres are separated by the distance $R$, in accord with the 
inference mentioned above. In the limit of $R\to0$ formula (\ref{Dq15}) 
yields
\begin{equation}\label{Dq16}
\mOmega^{(2)}(0)=\frac{5\alpha}{16} ,
\end{equation}
in agreement with (\ref{Dq6}). Based on relations (\ref{Dq10}), (\ref{Dq14}) 
and (\ref{Dq16}), expression (\ref{Bq18}) at $n=2$ describes the potential of 
interest. 

Substituting (\ref{Dq15}) into formulae (\ref{Dq1}) and (\ref{Dq2}), one can 
obtain the next generation of results appropriate to $n=3$. They are of the 
form
\begin{eqnarray}
\tW^{(3)}(z)=\Bigl(1+\frac{193z}{256}+\frac{65z^2}{256}+\frac{37z^3}{768}
+\frac{z^4}{192}+\frac{z^5}{3840}\Bigr)\exp(-z) ,\label{Dq17}\\
\mOmega^{(3)}(0)=\frac{63\alpha}{256} ,\label{Dq18}
\end{eqnarray}
where the transformation of $\mOmega^{(3)}(R)$ to $\tW^{(3)}(z)$ is carried
out with making use of (\ref{Bq19}) and (\ref{Dq11}) again. Starting from 
equations (\ref{Dq1}), (\ref{Dq2}) and (\ref{Dq17}), results for $n>3$ can be 
obtained in the same manner.

\section{Invariance of the Ewald approach}
Now we consider Gaussian functions of spreading. Interested in the spreading
of the $n$th order, we introduce
\begin{equation}\label{Eq1}
\sigma_n(r)=\Bigl(\frac{n\mu^2}{\pi}\Bigr)^{3/2}\exp\bigl(-n\mu^2r^2\bigr)
\end{equation}
that is normalized by condition (\ref{Bq2}). Substituting (\ref{Eq1}) into 
(\ref{Bq10}), integrating the result in Cartesian coordinates and keeping 
in mind the familiar Poisson integral \cite{Whit27}
\begin{equation}\label{Eq2}
\int_0^\infty\exp\bigl(-tu^2\bigr)\cos\bigl(qu\bigr)\,du
=\frac{1}{2}\sqrt{\frac{\pi}{t}}\exp\Bigl(-\frac{q^2}{4t}\Bigr) ,
\end{equation}
we get
\begin{equation}\label{Eq3}
S_n(\hb)=\exp\Bigl(-\frac{\pi^2|\hb|^2}{n\mu^2}\Bigr) .
\end{equation}
With making use of (\ref{Eq3}), relation (\ref{Bq14}) takes the form
\begin{equation}\label{Eq4}
U^{(n)}(\rb)=\frac{1}{\pi}\sump_\hb\frac{F(\hb)}{|\hb|^2}\exp
\Bigl(-\frac{\pi^2|\hb|^2}{\mu^2}+2\pi i\hb\rb\Bigr)
\end{equation}
that turns out to be independent of $n$.

The consideration of $\mOmega^{(n)}(R)$ in the particular cases of $n=1$
and $n=2$ may be performed basing on relations (\ref{Dq3}) and (\ref{Dq5}),
respectively. However, even at $n=2$ the corresponding relation is rather
complicated. It is evident that the complexity will further enhance for
$n>2$. This obstacle can be overcome within the approach proposed by Boys
\cite{Boys50}, where the integration over angular variables turns out to
be much more efficient if those variables are incorporated directly into
the exponents of Gaussian functions, as shown in \ref{App2}.

Let us consider, for a moment, the spreading function without normalization:
\begin{equation}\label{Eq5}
\sigma_\alpha(r)=\exp(-\alpha r^2) .
\end{equation}
According to results (\ref{Yq4}) and (\ref{Yq9}) from \ref{App2},
we then obtain
\begin{eqnarray}
\mOmega_\alpha^{(1)}(R)&=&\frac{2\pi}{\alpha}\frac{1}{R}\int_0^R
\exp(-\alpha r^2)\,dr ,\label{Eq6}\\
\mOmega_\alpha^{(2)}(R)&=&\frac{2\pi}{\alpha\sqrt{2}}\Bigl(
\frac{\pi}{\alpha}\Bigr)^{3/2}\frac{1}{R}\int_0^R
\exp\Bigl(-\frac{\alpha}{2} r^2\Bigr)\,dr .\label{Eq7}
\end{eqnarray}
Based on formulae (\ref{Eq6}), (\ref{Eq7}) and (\ref{Yq9}) from \ref{App2}, 
one can prove by induction that
\begin{equation}\label{Eq8}
\mOmega_\alpha^{(n)}(R)=\frac{2\pi}{\alpha\sqrt{n}}\Bigl[\Bigl(
\frac{\pi}{\alpha}\Bigr)^{3/2}\Bigr]^{n-1}\frac{1}{R}
\int_0^R\exp\Bigl(-\frac{\alpha}{n} r^2\Bigr)\,dr .
\end{equation}
Now in (\ref{Eq8}) we replace spreading function (\ref{Eq5}) 
by the normalized one described by (\ref{Eq1}). As a result, formula
(\ref{Eq8}) is transformed into
\begin{equation}\label{Eq9}
\mOmega^{(n)}(R)=\frac{2\mu}{\sqrt{\pi}R}\int_0^R\exp(-\mu^2 r^2)\,dr .
\end{equation}
It is clear that the dependence upon $n$ disappears here. In the
particular case of $R=0$ relation (\ref{Eq9}) yields
\begin{equation}\label{Eq10}
\mOmega^{(n)}(0)=\frac{2\mu}{\sqrt{\pi}} .
\end{equation}
On the other hand, relation (\ref{Eq9}) can be identically rewritten as:
\begin{equation}\label{Eq11}
\mOmega^{(n)}(R)=\frac{1}{R}-\frac{\erfc(\mu R)}{R} ,
\end{equation}
where the complementary error function
\begin{equation}\label{Eq12}
\erfc(z)=\frac{2}{\sqrt{\pi}}\int_z^\infty\exp(-u^2)\,du=\tW^{(n)}(z)
\end{equation}
just describes the value of $\tW^{(n)}(z)$ at $z=\mu R$ due to the last
equality that follows upon comparing (\ref{Eq11}) with (\ref{Bq19}) and
taking (\ref{Dq11}) into account.

If we substitute results (\ref{Eq4}), (\ref{Eq10}) and (\ref{Eq12}) into
(\ref{Bq18}), then we derive
\begin{eqnarray}\label{Eq13}
U(\rb)&=&\frac{1}{\pi}\sump_\hb\frac{F(\hb)}{|\hb|^2}\exp\Bigl(
-\frac{\pi^2|\hb|^2}{\mu^2}+2\pi i\hb\rb\Bigr)\nonumber\\
&&{}+\sump_i\int_V\frac{\rho(\rb')\erfc(\mu|\tRi|)\,d\rb'}{|\tRi|}
-\Bigl\{\frac{2\mu q_j}{\sqrt{\pi}}\Bigr\}_{\rb=\bb_j} ,
\end{eqnarray}
where $\tRi$ is defined by (\ref{Aq16}). On making use of formula
(\ref{Eq13}) in (\ref{Cq13}), the specific energy takes the form
\begin{eqnarray}\label{Eq14}
\Ec&=&\frac{v}{2\pi}\sump_\hb\frac{|F(\hb)|^2}{|\hb|^2}
\exp\Bigl(-\frac{\pi^2|\hb|^2}{\mu^2}\Bigr)\nonumber\\
&&+\frac{1}{2}\sump_i\int_V \frac{\rho(\rb)\rho(\rb')
\erfc(\mu|\tRi|)}{|\tRi|}d\rb\,d\rb'-\frac{\mu}{\sqrt{\pi}}\sum_j q_j^2 .
\end{eqnarray}
Equations (\ref{Eq13}) and (\ref{Eq14}) are the classical formulae of
Ewald \cite{Ewal21}. We draw a conclusion that Gaussian spreading
functions appear to be invariant with respect to their multiple
application, without changing the functional form of the result.

\section{Multiple charge spreading in individual pair interactions}
According to (\ref{Bq17}), the potential effect of spreading charges is 
separated from that of the initial ones in crystals. Therefore the final 
analytical results depend solely on a single value of $n$ specifying the 
order of a multiple spreading at hand. The rate of convergence in dependence 
on $n$ is a special subject that will be discussed elsewhere \cite{Kho208}.

Here we concentrate our attention on another manifestation of charge
spreading keeping in mind that in a single neutral object the charge
spreading distribution may be treated as neutralizing a more compact charge
of opposite sign \cite{Whee02}. Let us consider two complex objects of this 
sort, which are in general defined by the charge distributions
\begin{eqnarray}
\rho_1(\rb)=Z_1\Bigl[f^{(m_1)}(\Zeb,\rb)-f^{(n_1)}(\Zeb,\rb)\Bigr] ,
\label{Fq1}\\
\rho_2(\rb)=Z_2\Bigl[f^{(m_2)}(\Zeb,\rb)-f^{(n_2)}(\Zeb,\rb)\Bigr] ,
\label{Fq2}
\end{eqnarray}
where $Z_1$ and $Z_2$ are the total charges describing either part of 
$\rho_1(\rb)$ and $\rho_2(\rb)$, respectively, in accord with (\ref{Cq4}) 
and (\ref{Cq6}). Here we restrict ourselves to a multiple
application of a certain initial spreading function so that relations 
(\ref{Cq1})--(\ref{Cq3}) are taken into account in definitions (\ref{Fq1}) 
and (\ref{Fq2}). The energy of interaction between charge densities 
(\ref{Fq1}) and (\ref{Fq2}) separated by the distance $R$ can be written 
in the conventional form as
\begin{equation}\label{Fq3}
E(R)=\int\frac{\rho_1(\rb_1)\rho_2(\rb_2)\,d\rb_1d\rb_2}{|\Rb+\rb_1-\rb_2|} .
\end{equation}
On substituting (\ref{Fq1}) and (\ref{Fq2}) into (\ref{Fq3}) and taking 
formula (\ref{Cq10}) into account, expression (\ref{Fq3}) can be transformed 
into
\begin{eqnarray}\label{Fq4}
E(R)&=&Z_1Z_2\Bigl[\mOmega^{(m_1+m_2)}(R)-\mOmega^{(m_1+n_2)}(R)
-\mOmega^{(n_1+m_2)}(R)\nonumber\\
&&+\mOmega^{(n_1+n_2)}(R)\Bigr]=\frac{Z_1Z_2}{R}\Bigl[W^{(m_1+n_2)}(R)
\nonumber\\
&&+W^{(n_1+m_2)}(R)-W^{(m_1+m_2)}(R)-W^{(n_1+n_2)}(R)\Bigr] .
\end{eqnarray}
Here each $\mOmega^{(\dots)}(R)$ describes the interaction energy between
the corresponding single terms in (\ref{Fq1}) and (\ref{Fq2}) and the
transition to the last relation is performed by means of (\ref{Bq19}). As a 
result, four different orders of spreading appear in this general case.
Of course, a particular event at $m_1=m_2=0$ and $n_1=n_2=1$ is of special
interest. In this case $W^0(R)=0$ due to (\ref{Bq19}), (\ref{Cq2}) and 
(\ref{Cq10}). Formula (\ref{Fq4}) is then converted into
\begin{equation}\label{Fq5}
E(R)=\frac{Z_1Z_2}{R}\Bigl[2W^{(1)}(R)-W^{(2)}(R)\Bigr] ,
\end{equation}
where only two consecutive orders of spreading happen.

The particular case associated with a simple exponential spreading arises
after inserting relations (\ref{Dq12}) and (\ref{Dq14}) into (\ref{Fq5}).
Keeping (\ref{Dq11}) in mind, we then obtain
\begin{equation}\label{Fq6}
E_{\rm s}(z)=\frac{\alpha Z_1Z_2}{z}\Bigl[1+\frac{5z}{16}-\frac{3z^2}{16}
-\frac{z^3}{48}\Bigr]\exp(-z) ,
\end{equation}
where $z=\alpha R$. Likewise, the case appropriate to Gaussian spreading
functions arises upon substituting  (\ref{Eq12}) into (\ref{Fq5}), with 
taking underlying definition (\ref{Eq1}) into account. Starting from 
$\sigma_1(r)$ in (\ref{Eq1}), we readily obtain the following result 
\begin{equation}\label{Fq7}
E_{\rm G}(z)=\frac{\mu Z_1Z_2}{z}\Bigl[2\erfc(z)-\erfc\bigl(z/\sqrt{2}
\bigr)\Bigr] ,
\end{equation}
where $z=\mu R$ and $\mu^2\to2(\mu/\sqrt{2})^2$ in definition (\ref{Eq1}) 
so as to describe the last term in the parentheses in (\ref{Fq7}).

Comparing issues (\ref{Fq6}) and (\ref{Fq7}), we recognize one more 
universality, which now corresponds to the energy described by the simple
exponential spreading, where the same exponent turns out to be typical of
both the terms addressed to $W^{(1)}(R)$ and $W^{(2)}(R)$. Actually, it is
\begin{figure}[t]
\begin{center}
\resizebox{0.5\hsize}{!}{\includegraphics{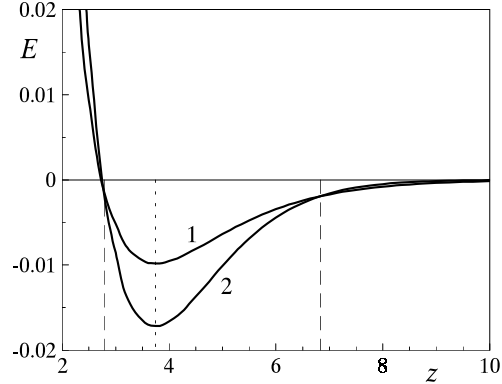}}
\end{center}
\caption{The "interatomic" energy $E$, in units of $\alpha Z_1Z_2$, versus 
a dimensionless distance $z=\alpha R$. Curve 1 is appropriate to equation 
(\ref{Fq6}) and exhibits the case of a simple exponential charge clouds 
neutralizing central point charges of the distance $R$ apart. The case of 
the corresponding Gaussian charge clouds specified by formula (\ref{Fq7}) 
is described by curve 2, providing that its minimum coincides with the
minimum point of curve 1 at $\mu=0.3586\alpha$ that is marked by the vertical
dotted line. The dashed lines indicate the points of intersection of these 
curves.}\label{Fig1}
\end{figure}
not surprising because the next sample of this set, i.e. $W^{(3)}(R)$
described by (\ref{Dq17}), is specified by the same exponential decrease. 
Conversely, if the energy is determined by Gaussian spreading functions, 
then either of the contributions to (\ref{Fq7}) is specified by its own 
predominant law of decrease.

It seems to be interesting to recall that there is sometimes a tendency
towards changing a simple exponential electron spreading by a Gaussian one
in molecular calculations, where the contribution of the exchange interaction
can then be evaluated in a much simpler manner \cite{Boys50,Ween53,Lomb66,%
Cast98}. Although the direct Coulomb interaction is still the subject of our 
interest, this is the reason to compare results (\ref{Fq6}) and (\ref{Fq7}) 
in more amount of detail. To this end, we plot the corresponding energy 
curves together, as shown in figure \ref{Fig1}. The shape of either of these 
curves is quite natural. Indeed, the energy is positive and its value tends to 
infinity as $R$ drops to zero. On the other hand, if $R$ grows, then the energy 
eventually becomes negative, attains at its minimum value and farther falls off
\begin{figure}[t]
\begin{center}
\resizebox{0.5\hsize}{!}{\includegraphics{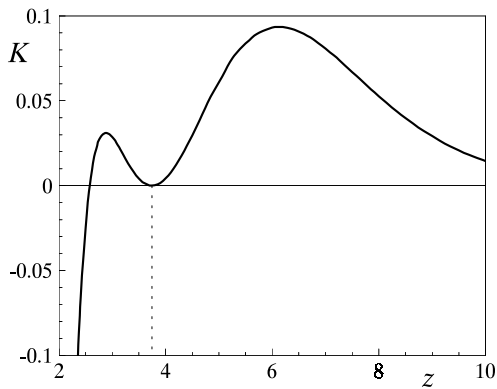}}
\end{center}
\caption{The parameter $K$ defined by formula (\ref{Fq8}) versus a
dimensionless distance $z=\alpha R$. The position of energy minima,
where $K$ has a local minimum, is pointed out by the vertical dotted 
line.}\label{Fig2}
\end{figure}
to zero in magnitude, being still negative. The latter is a direct consequence
of the fact that just in general relation (\ref{Fq5}) the contribution of
$W^{(2)}(R)$ as a function of $R$ is always more diffuse than that of 
$W^{(1)}(R)$. In order to compare both the curves, we have shifted a minimum 
point of the curve describing quantity (\ref{Fq7}) to the value of 
$R=3.745/\alpha$. We see that a minimum of curve 2 corresponding to 
$E_{\rm G}^{\rm{min}}=-0.01715\alpha Z_1Z_2$ is much deeper than that of 
curve 1 with the value of $E_{\rm s}^{\rm{min}}=-0.009805\alpha Z_1Z_2$ and 
this effect is described by the ratio 1.749. Moreover, the curvature of 
curve 2 at its minimum point is also greater than that of curve 1. As a 
result, there are two points of intersection between those curves which take 
place at $R=2.781/\alpha$ and at $R= 6.830/\alpha$, with the energy values 
$E=-0.000652\alpha Z_1Z_2$ and $E=-0.00194\alpha Z_1Z_2$, respectively, as 
shown in figure \ref{Fig1} as well. 

Of course, the energy $E_{\rm G}(z)$ can be further scaled by the factor
$E_{\rm s}^{\rm{min}}/E_{\rm G}^{\rm{min}}$ so as to simulate the behaviour
of $E_{\rm s}(z)$. The comparison of both these energies is then specified by 
the relative value of the form
\begin{equation}\label{Fq8}
K(z)=\Bigl[E_{\rm G}(z)\frac{E_{\rm s}^{\rm{min}}}{E_{\rm G}^{\rm{min}}}
-E_{\rm s}(z)\Bigr]\frac{1}{|E_{\rm s}^{\rm{min}}|}
=\frac{E_{\rm G}(z)}{|E_{\rm G}^{\rm{min}}|}
-\frac{E_{\rm s}(z)}{|E_{\rm s}^{\rm{min}}|}.
\end{equation}
The behaviour of $K(z)$ as a function of $z$ is shown in figure \ref{Fig2}. We
see that $K(z)$ turns out not to be monotonic in the vicinity of $z$ corresponding
to minima of the energies at hand. This fact can be important upon minimizing 
a total energy modified by other energy contributions. In this case the effect
driven by simple exponential spreading functions and that driven by Gaussian
spreading functions are expected to be far from being proportional.

\section{Conclusion}
In summary, it is shown that the effect of charge spreading proposed by Bertaut 
\cite{Bert52,Bert78} can be utilized in a multiple manner. It means that the 
problem how many times a given spreading function is applied to the original 
charge distribution in a crystal is not of principle. Nevertheless, the tendency 
towards increasing the rate of convergence upon multiple charge spreading is 
just recognized by Bertaut \cite{Bert78} and will be confirmed elsewhere
\cite{Kho208}. This result is not trivial. Presumably, it is associated with 
an idea that there is an optimum spreading configuration with very diffuse 
tails. In this connection, the fact that all the effects driven by a Gaussian 
spreading function are reproduced in the same functional form, regardless of its 
multiple application, may anyhow point to an optimum character of a Gaussian 
spreading.

Here we also recognize that a certain spreading, either single or multiple, 
may be attributed to every point of space and so may be treated as a uniform 
transformation of space. It is significant that the general relation between 
electrostatic potentials and specific Coulomb energies in crystals, as well as 
zero value of the mean potential there, turns out to be invariant with respect 
to such a transformation. 

It is evident that the application of a multiple charge spreading to the problem
of lattice summation is nothing but a fruitful approach to that problem. It 
implies that the final results of lattice summation are to be independent of 
the shape of spreading. However, it is not the case if the charge spreading is
regarded as a real property of at least a pair of complex neutral physical 
objects connected by the Coulomb interaction. In this event the replacement of 
a natural shape of, for example, an electron cloud with a more artificial shape 
must be performed with caution.

\appendix
\section{Some general relations for charge spreading}\label{App1}
Equation (\ref{Bq16}) can be readily rewritten as
\begin{equation}\label{Zq1}
\mOmega^{(n)}(|\Rb|)=\int\sigma(|\rb|)\mOmega^{(n-1)}(|\Rb+\rb|)
\,d\rb .
\end{equation}
With making use of spherical coordinates of $\rb$, we assume that
\begin{equation}\label{Zq2}
|\Rb+\rb|=\bigl(R^2+r^2+2Rr\cos\theta\bigr)^{1/2}\equiv y
\end{equation}
and go over from the variable $\theta$ to a new variable $y$. The result 
of integration over $y$ is then of form (\ref{Dq1}). On the other hand,
if $\Rb=0$, then the integration over angular variables of $\rb$ in
(\ref{Zq1}) is trivial and we obtain issue (\ref{Dq2}).

The case of $\mOmega^{(1)}(R)$ is straightforward and is described by 
\begin{equation}\label{Zq3}
\mOmega^{(1)}(R)=\frac{1}{R}-4\pi\int_R^\infty\sigma(r)\Bigl(\frac{r}{R}
-1\Bigr)r\,dr .
\end{equation}
Equations (\ref{Dq3}) and (\ref{Dq4}) follow therefrom. If we are interested 
in $\mOmega^{(2)}(R)$, then the employment of (\ref{Zq3}) in the general 
relation (\ref{Dq1}) gives rise to
\begin{equation}\label{Zq4}
\mOmega^{(2)}(R)=\mOmega^{(1)}(R)-\frac{8\pi^2}{R}\int_0^\infty
\sigma(r)J(R,r)r\,dr ,
\end{equation}
where we utilized definition (\ref{Zq3}) again and
\begin{equation}\label{Zq5}
J(R,r)=\int_{|R-r|}^{R+r}dy\int_y^\infty\sigma(r')(r'-y)r'\,dr' .
\end{equation}
If we interchange the order of integration over $r'$ and $y$ here, 
then the integration over $y$ is straightforward and we obtain
\begin{equation}\label{Zq6}
\fl J(R,r)=\frac{1}{2}\int_{|R-r|}^\infty\sigma(r')\bigl(|R-r|-r'\bigr)^2
r'\,dr'-\frac{1}{2}\int_{R+r}^\infty\sigma(r')\bigl(R+r-r'\bigr)^2r'\,dr' .
\end{equation}
In turn, based on (\ref{Bq2}), it is expedient to rewrite expression
(\ref{Zq3}) in an identical form
\begin{equation}\label{Zq7}
\mOmega^{(1)}(R)=\frac{1}{R}-\frac{16\pi^2}{R}\int_0^\infty
\sigma(r)r\,dr\int_R^\infty\sigma(r')r'\bigl[r(r'-R)\bigr]dr' .
\end{equation}
Substituting (\ref{Zq6}) and (\ref{Zq7}) into (\ref{Zq4}) and
combining the integral terms, we arrive at the result in the most
symmetric form given by formulae (\ref{Dq5}), (\ref{Dq7}) and (\ref{Dq8}),
in agreement with (\ref{Bq19}). On the other hand, relation (\ref{Dq6}) 
for $\mOmega^{(2)}(0)$ appears directly upon substituting (\ref{Zq3}) 
into (\ref{Dq2}).

\section{Coulomb interaction between Gaussian functions}\label{App2}
Here we follow the treatment of Boys \cite{Boys50}. Let us consider two
charge distributions
\begin{equation}\label{Yq1}
\sigma_\alpha(r)=\exp(-\alpha r^2) ,\quad
\sigma_\beta(r)=\exp(-\beta r^2) .
\end{equation}
According to (\ref{Bq16}), they determine the values
\begin{eqnarray}
&&\mOmega^{(1)}_{\beta}(R)=\int\frac{\sigma_\beta(r)\,d\rb}{|\Rb
+\rb|} ,\label{Yq2}\\
&&\mOmega^{(2)}_{\alpha\beta}(R)=\int\sigma_\alpha(r)
\mOmega^{(1)}_{\beta}(|\Rb+\rb|)\,d\rb ,\label{Yq3}
\end{eqnarray}
where $R=|\Rb|$ and $r=|\rb|$.

The particular form of the distributions in (\ref{Yq1}) enables one to
go over to the variable $\rb'=\rb+\Rb$ in equation (\ref{Yq2}). The 
integration over the angular coordinates of $\rb'$ is then trivial there 
and we obtain
\begin{equation}\label{Yq4}
\mOmega^{(1)}_{\beta}(R)=\frac{2\pi}{\beta R}\int_0^R\exp(-\beta x^2)\,dx,
\end{equation}
where $x=|\rb'|$. Inserting (\ref{Yq4}) into (\ref{Yq3}) and operating 
further in the same manner, we derive
\begin{equation}\label{Yq5}
\mOmega^{(2)}_{\alpha\beta}(R)=\frac{2\pi^2}{\alpha\beta R}I(R) ,
\end{equation}
where
\begin{equation}\label{Yq6}
I(R)=\int_{-\infty}^\infty\exp[-\alpha(R-y)^2]\,dy\int_0^y 
\exp(\beta x^2)\,dx.
\end{equation}
If we differentiate equation (\ref{Yq6}) with respect to $R$, then we obtain
\begin{equation}\label{Yq7}
\frac{dI(R)}{dR}=-\int_{-\infty}^\infty\frac{d}{dy}\Bigl\{
\exp[-\alpha(R-y)^2]\Bigr\}\,dy\int_0^y\exp(\beta x^2)\, dx .
\end{equation}
Integrating the right-hand side of (\ref{Yq7}) by parts, we easily reach
\begin{equation}\label{Yq8}
\frac{dI(R)}{dR}=\sqrt{\frac{\pi}{\alpha+\beta}}\exp\Bigl(
-\frac{\alpha\beta}{\alpha+\beta}R^2\Bigr) .
\end{equation}
Note that $I(0)=0$ follows from (\ref{Yq6}) and specifies the further
integration of (\ref{Yq8}) with respect to $R$. On inserting the result 
of integration into equation (\ref{Yq5}), the final issue takes the form
\begin{equation}\label{Yq9}
\mOmega^{(2)}_{\alpha\beta}(R)=\frac{2\pi^{5/2}}{\alpha\beta
\sqrt{\alpha+\beta}R}\int_0^R\exp\Bigl(-\frac{\alpha\beta z^2}{\alpha
+\beta}\Bigr)\,dz
\end{equation}
that is naturally symmetric with respect to $\alpha$ and $\beta$.

\end{document}